\documentclass[aps,prb,twocolumn,groupedaddress]{revtex4}

\usepackage{graphicx}
\usepackage{dcolumn}

\begin{document}


\title{Hole correlation and antiferromagnetic order in the
  $t$-$J$ model 
} 

\author{P. W. Leung}
\email[]{P.W.Leung@ust.hk}
\affiliation{Physics Dept., Hong Kong University of Science and Technology,
Clear Water Bay, Hong Kong}

\date{\today}

\begin{abstract}
We study the $t$-$J$ model with four holes on a 32-site square lattice
using exact diagonalization. This system corresponds to doping level $x=1/8$.
At the ``realistic'' parameter $J/t=0.3$, holes in the ground state of
this system are 
unbound. They have
short range repulsion due to lowering of kinetic energy. There is no
antiferromagnetic spin order and the electron momentum 
distribution function resembles hole
pockets. 
Furthermore, we show evidence that in case antiferromagnetic order
exists, holes form $d$-wave
bound pairs and there is mutual repulsion among hole pairs. This
presumably will occur at low doping level. This scenario is
compatible with a checkerboard-type charge density state proposed to
explain the ``1/8 
anomaly'' in the LSCO family, except that it is the ground state only
when the system possesses strong antiferromagnetic order.

\end{abstract}

\pacs{
71.27.+a, 
71.10.Fd, 
75.40.Mg 
}

\maketitle

It is generally accepted that the physics of high temperature
superconductors is the physics of a doped antiferromagnetic
insulator. Although the detail mechanism is still not fully
understood, it is believed that frustration due to mobile holes may lead to
superconductivity as well as other competing orders\cite{sz02,kbf03,lee04} 
which may include
charge ordered and spin ordered phases, 
or even more exotic phases like the
staggered-flux phase. One way to examine the possibilities of these
states is to start from a lightly doped antiferromagnet.
The $t$-$J$ model with two holes has been extensively studied using
various analytical and numerical methods.\cite{e92,ks93,p94,bcs97,clg98,l02}
These results indicate
that two holes form a weak bound pair with $d_{x^2-y^2}$
symmetry.
In addition, exact diagonalization study of the two-hole model
lends support to the staggered-flux phase.\cite{l00}
Nevertheless there are still many open questions.
In a system with two holes the electron
momentum distribution function (EMDF) $\langle n_{{\bf k}\sigma}\rangle$
reflects the structure of the bound pair only and may
not contain information on the Fermi surface in the thermodynamic
limit. 
A system with four holes is therefore the first step in possibly
extracting useful information on the system at finite doping level.
Furthermore, 
the two-hole system has a few low-lying excited states where the holes
do not form a $d_{x^2-y^2}$ bound pair.\cite{l02}
It is
therefore unclear whether $d$-wave hole pairing still
survives when more holes are present. If so, how would hole pairs
interact? And if not, what scenario replaces hole pairing?

\begin{figure}
 \resizebox{8cm}{!}{\includegraphics{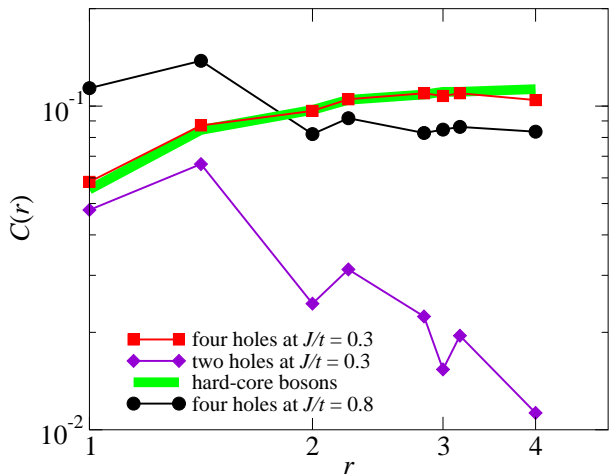}}
 \caption{\label{fig:cr} (Color online) Hole-hole correlation in the
   ground states 
   with two
   and four holes. The thick shaded line is the pair correlation
   function of four hard 
   core bosons on the same lattice. }
 \end{figure}

Motivated by  the above questions, we numerically diagonalize the
$t$-$J$ model with
four holes on a
32-site lattice with periodic boundary conditions. We caution that due
to the small size of our system we cannot study long-range
orders. Therefore superconductivity is not the subject of this study.
The
doping level of this system is $x=1/8$ which is in the underdoped
region. Incidentally this is the same doping level as the
``1/8 anomaly'' in the LSCO family
where experimental signals for 
charge ordering are more pronounced.\cite{kbf03}
The Hamiltonian of the $t$-$J$ model is 
\begin{equation}
\label{tJ}
{\cal H}=-t\sum_{\langle ij\rangle\sigma} (\tilde{c}^\dagger_{i\sigma}
\tilde{c}_{j\sigma}+h.c.)
+J\sum_{\langle ij\rangle} \left({\bf S}_i\cdot{\bf S}_j
-\frac{n_in_j}{4}\right),
\end{equation}
where $\langle ij\rangle$ denotes a nearest neighbor pair. 
Let us consider the case
$J/t=0.3$ first. This parameter is believed to be relevant
to high $T_c$ cuprates.
The ground state has energy
$-18.682017t$, momentum
$(0,0)$ and 
$d_{x^2-y^2}$ symmetry. (Note that this refers to the symmetry of the
total wave function. It is not the pairing symmetry, nor does it imply
hole pairing.)
To see whether there is hole binding we calculate the
hole-hole correlation 
function\cite{cr}
$C(r)=\langle (1-n_r)(1-n_0)\rangle$,
where $n_r$ is the spin number
operator in Eq.~(\ref{tJ}). 
Fig.~\ref{fig:cr} shows that at $J/t=0.3$, systems with two and four
holes have completely different behaviors. While holes in the two-hole
system form a bound pair, in the four-hole system they are
mutually repulsive
--- $C(r)$ increases with $r$ and does not vary much beyond $r=2$.
The root-mean-square separation between holes
is $r_{\text{rms}}\equiv\sqrt{\langle r^2\rangle}=2.4732$. This is
slightly larger than  2.3827, which is the root-mean-square separation
between uncorrelated holes on the same lattice.
In order to understand this repulsion, we compare
$C(r)$ to 
the pair correlation function
of four hard core bosons on the same lattice.
Besides the condition of no double occupancy, hard
core bosons have effective short-range repulsion due to hopping,
but are otherwise uncorrelated.
As shown in Fig.~\ref{fig:cr}, the pair correlation function of hard
core bosons is strikingly similar to $C(r)$ of our four-hole
system. $r_{\text{rms}}$ of the bosons is $2.4894$, which is 
very close to that of the four-hole system. 
From this similarity we conclude that in
the $t$-$J$ model with four holes at $J/t=0.3$, holes have short-range
repulsion which arises from
kinetic effect due to hopping.

\begin{figure}
   \resizebox{8cm}{!}{\includegraphics{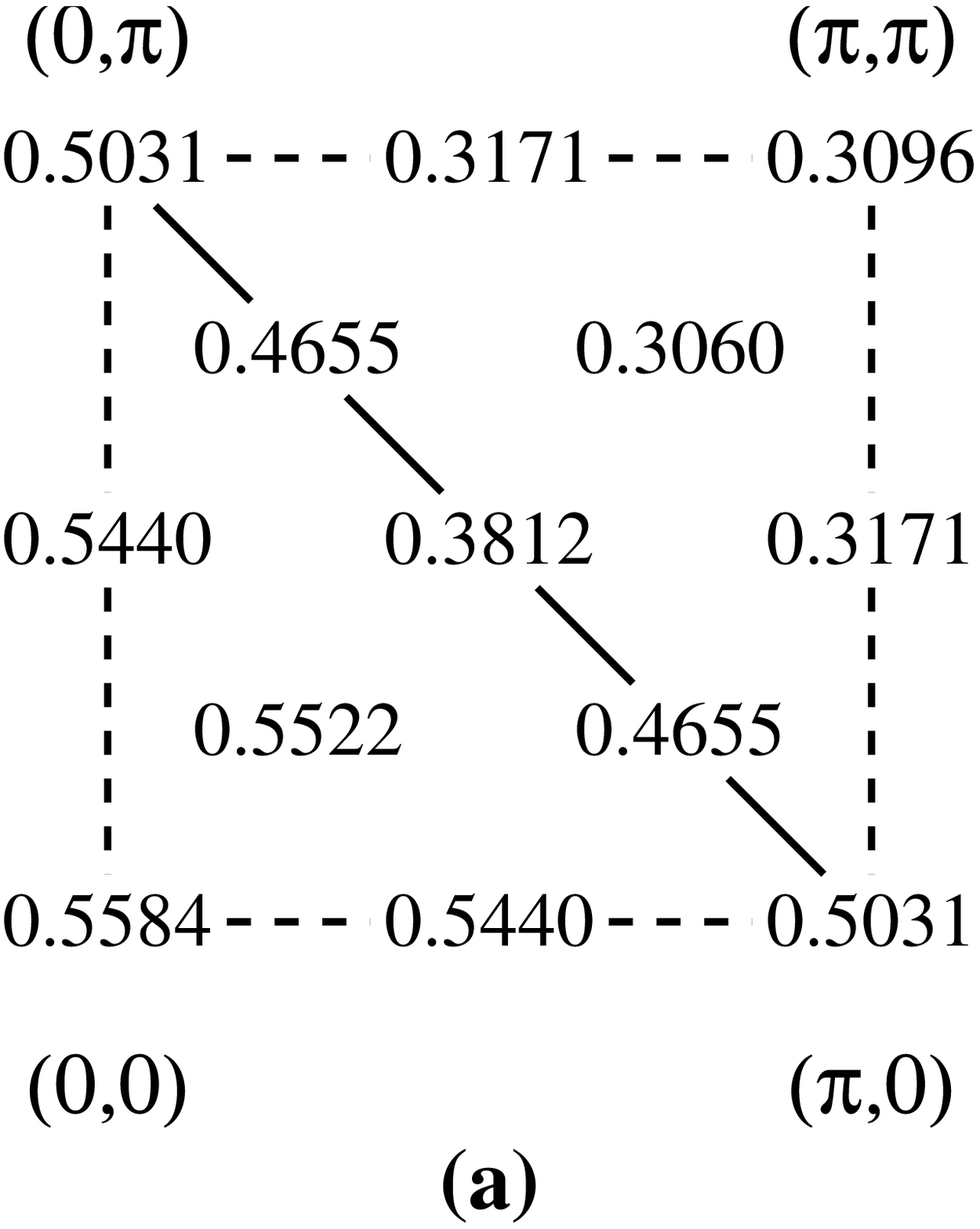}\hspace{1cm}
   \includegraphics{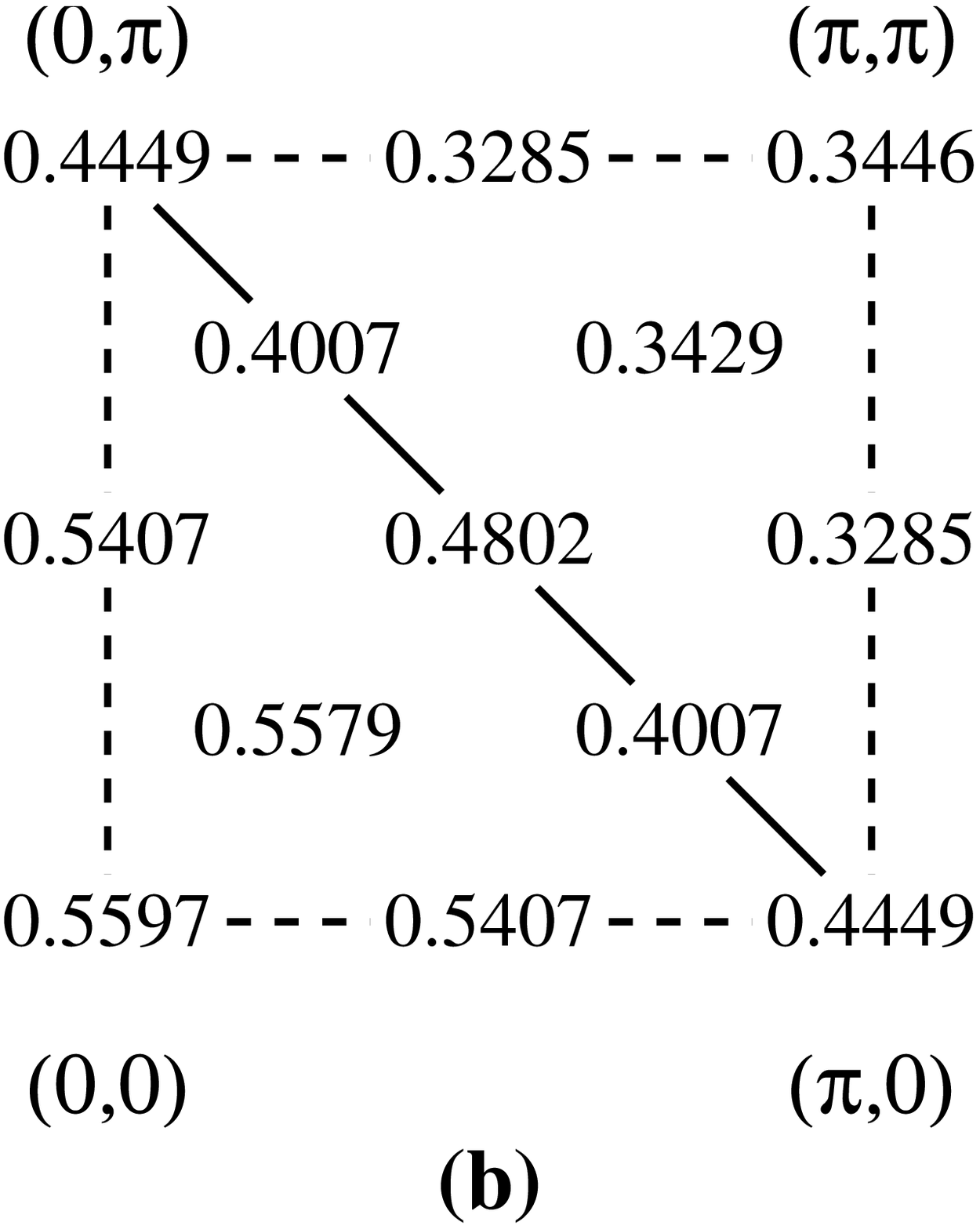}}  
 \caption{\label{fig:h_nq} EMDF $\langle n_{\bf
   k}\rangle$ in the ground state of the four-hole system at 
   (a) $J/t=0.3$ and (b) $J/t=0.8$. Solid line is the AFBZ
   boundary. Due to symmetry, only one
   quadrant of the Brillouin zone is shown.}
\end{figure}

More information on the interaction among holes comes from correlation
in momentum space. As usual, we define the electron momentum
distribution function (EMDF)
as $\langle n_{{\bf k}\sigma}\rangle
\equiv\langle \tilde{c}^\dagger_{{\bf k}\sigma}\tilde{c}_{{\bf
k}\sigma}\rangle$. 
The result is shown in Fig.~\ref{fig:h_nq}(a). 
Since the electron number is even, $\langle n_{{\bf
    k}\uparrow}\rangle=\langle n_{{\bf k}\downarrow}\rangle$ and we
drop the spin variable $\sigma$ from $\langle n_{{\bf
    k}\sigma}\rangle$.
In interpreting the result it is important to realize that the general
shape of the EMDF can be misleading.
It features a ``dome''-shape
with a maximum at $(0,0)$ and slopes down towards
$(\pi,\pi)$. 
This is a general feature at all doping levels and
has been shown\cite{ew93,eo95} to result from minimizing the kinetic
energy. It does not represent the true Fermi
surface. 
Instead we should focus on those {\bf k} points along the boundary of
the antiferromagnetic Brillouin zone (AFBZ), i.e., from $(\pi,0)$ to
$(0,\pi)$. They are not affected by 
the kinematic effect and therefore reflect the physics of the
system.\cite{clg98}
Fig.~\ref{fig:h_nq}(a) shows that along the AFBZ boundary the EMDF has a
minimum at $(\pi/2,\pi/2)$. 
Again this is in sharp contrast with the two-hole system
whose EMDF has a maximum at $(\pi/2,\pi/2)$ along
the AFBZ boundary.\cite{clg98}
Unfortunately we do not
have useful information on how the EMDF changes from
$(0,0)$ to $(\pi,\pi)$ because kinematic effect dominates
along this direction. Therefore we cannot conclude whether the
minimum at $(\pi/2,\pi/2)$ implies a
hole pocket or a half pocket (segmented Fermi surface). But 
it is not in contradiction with either one. 

The fact that the EMDF of the four-hole system has a minimum at the
single-hole ground state momentum $(\pi/2,\pi/2)$ immediately suggests
the relevance of the single-hole ground state to the multiple-hole
one.\cite{sh90,eos94,eo95} 
If holes doped into the parent system behave like
weakly interacting fermions, then it is reasonable to expect that
many-hole systems 
can be approximated by filling up the single-hole band. This 
should lead to ``hole-pockets'' 
at single-hole ground state momenta.\cite{fulde}
We would like to see how well the EMDF 
of the four-hole system agrees with this simple 
picture. 
For this purpose we consider the hole 
momentum distribution function (HMDF),
\begin{equation}
\langle {\overline n}_{{\bf k}\sigma}\rangle
\equiv \langle\tilde{c}_{{\bf
    k}\sigma}\tilde{c}^\dagger_{{\bf k}\sigma}\rangle
= (N_\sigma+N_h)/N- \langle n_{{\bf
    k}\sigma}\rangle,
\end{equation}
where $N_\sigma$ and $N_h$ are numbers of $\sigma$ spins and holes
respectively. 
If multiple-hole states can be built up from single-hole states, we
expect their HMDFs to be additive,
\begin{eqnarray}
\langle{\overline n}_{\bf k}\rangle_2 &\simeq& \langle{\overline n}_{{\bf
    k}\uparrow}\rangle_1 + \langle{\overline n}_{{\bf
    k}\downarrow}\rangle_1,\label{eq:rigidband2}\\
\langle{\overline n}_{\bf k}\rangle_4 &\simeq& 2(\langle{\overline n}_{{\bf
    k}\uparrow}\rangle_1 + \langle{\overline n}_{{\bf
    k}\downarrow}\rangle_1),
\label{eq:rigidband4}
\end{eqnarray}
where $\langle{\overline n}_{{\bf k}\sigma}\rangle_{N_h}$ is the HMDF
of the $N_h$-hole system.
Table~\ref{tab:nq} shows the HMDF in the
two- and four-hole systems together with the approximations
from Eq.~(\ref{eq:rigidband2}) and (\ref{eq:rigidband4})
respectively.\cite{rb}
Note that Eq.~(\ref{eq:rigidband2}) and (\ref{eq:rigidband4}) lead to
very prominent 
maxima at $(\pi/2,\pi/2)$ 
along the AFBZ boundary. This of course does not agree with
our exact result in the two-hole system. Even in the
four-hole system, the maximum
is too sharp compared to the exact result.
Therefore a simple additive model as given
in Eq.~(\ref{eq:rigidband2}) and (\ref{eq:rigidband4}) does not
work in the $t$-$J$ model. We remark that such simple additive
picture holds in the extended $t$-$J$ model with parameters chosen to
represent electron-doped cuprates.\cite{edope}
We conclude that although the EMDF of the four-hole system exhibit
minima that suggests
the relevance of the single-hole ground state to the multiple-hole one,
there must still be appreciable
correlation among the holes.

\begin{table}
\caption{\label{tab:nq} HMDF $\langle{\overline n}_{\bf
  k}\rangle$ in the two- and four-hole systems 
  at $J/t=0.3$. {\bf k} points along the AFBZ boundary are marked with
  $\dagger$. Numbers in 
  brackets are calculated using the additive assumption,
  Eq.~(\ref{eq:rigidband2}) and (\ref{eq:rigidband4}), where
  $\langle{\overline n}_{\bf k}\rangle_1$ are from Ref.~\onlinecite{clg98}.}
\begin{ruledtabular}
 \begin{tabular}{ccc}
{\bf k}&
two-hole system&
four-hole system\\
\colrule
 (0,0)                                  &  0.0031  (0.0058)&  0.0041  (0.0115)\\
 ($\frac{\pi}{4}$,$\frac{\pi}{4}$)      &  0.0042  (0.0064)&  0.0103  (0.0128)\\
 ($\frac{\pi}{2}$,$\frac{\pi}{2}$)$^\dagger$      &  0.0376  (0.1156)&  0.1813  (0.2311)\\
 ($\frac{3\pi}{4}$,$\frac{3\pi}{4}$)    &  0.1234  (0.1416)&  0.2565  (0.2832)\\
 ($\pi$,$\pi$)                          &  0.1255  (0.1316)&  0.2529  (0.2633)\\
 ($\pi$,$\frac{\pi}{2}$)                &  0.1360  (0.0582)&  0.2454  (0.1164)\\
 ($\pi$,0)$^\dagger$                              &  0.0497  (0.0137)&  0.0594  (0.0274)\\
 ($\frac{\pi}{2}$,0)                    &  0.0104  (0.0021)&  0.0185  (0.0042)\\
 ($\frac{3\pi}{4}$,$\frac{\pi}{4}$)$^\dagger$     &  0.0657  (0.0169)&  0.0970  (0.0339)\\
\end{tabular}
\end{ruledtabular}
\end{table}

\begin{figure}
   \resizebox{8cm}{!}{\includegraphics{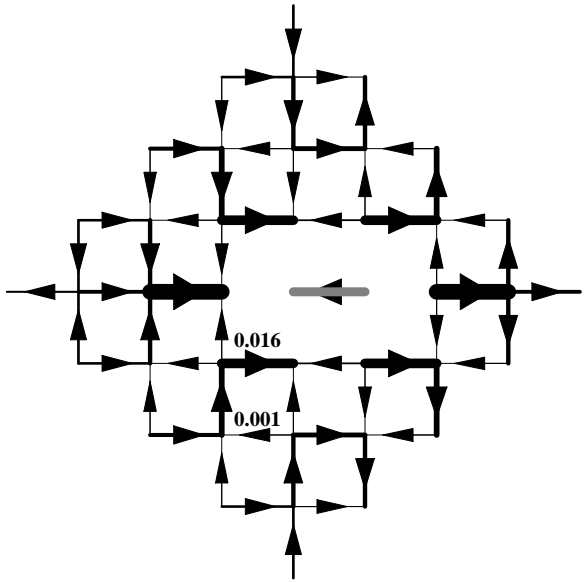}}   
  \caption{\label{fig:jj} Current correlation in the ground state of
    the four-hole system  
   at $J/t=0.3$. The
   reference bond $mn$ is indicated by a shaded line. Except for the
   reference bond, line widths are proportional to $\langle
   j_{kl}j_{mn}\rangle/x$, where $x=1/8$ is the doping level. For
   reference purpose, the numerical values of $\langle 
   j_{kl}j_{mn}\rangle/x$ are shown next to two horizontal bonds.  }
\end{figure}
\begin{figure}
   \resizebox{6cm}{!}{\includegraphics{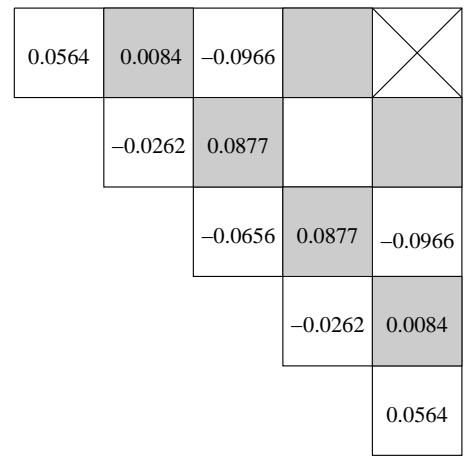}}
  \caption{\label{fig:vv} The vorticity correlation $C_{VV}(r)$ in the
    ground state of 
    the four-hole system at 
    $J/t=0.3$.
  The reference plaquette at $r=0$ is
  indicated by a cross inside it. 
  Because of symmetry, only a quarter of
  the cluster is  shown. }
\end{figure}

Next we look at a subtle form of correlation that is believed to
reflect hole binding.
A staggered pattern in the current correlation function $\langle
j_{kl}j_{mn}\rangle$, where 
$
j_{kl}=it(\tilde{c}^\dagger_k\tilde{c}_l-\tilde{c}^\dagger_l\tilde{c}_k)
$
is the hole current on the nearest neighbor bond $kl$, was first suggested as a
manifestation of the staggered flux phase.\cite{ilw00} Although there
exists another explanation in terms of the spin
polaron theory,\cite{we01} both theories point to the fact that such
staggered correlation reflects hole binding in the $d$-wave
channel. 
The current correlation at $J/t=0.3$ is shown in
Fig.~\ref{fig:jj}. We note that the direction of the correlation
function on many bonds deviate from a staggered pattern.
It certainly does not have the clear staggered 
pattern that exists in the two-hole system.\cite{l00} 
Another way to display the overall pattern is through
the vorticity correlation,
$C_{VV}(r)\equiv\langle V(r)V(0)\rangle/x$. The vorticity
$V({\bf r})$ of a square plaquette centered at {\bf r} is defined by
summing up the current correlations around it in the counterclockwise
direction.\cite{ilw00} We plot $C_{VV}(r)$ in Fig.~\ref{fig:vv}.
It clearly shows that the vorticity correlation is weak and  
decays rapidly, leading to deviation from a staggered pattern at large
distance. Hence from the current correlation we conclude that
there is no sign of $d$-wave pairing in the four-hole system at $J/t=0.3$.

\begin{figure}
    \resizebox{8cm}{!}{\includegraphics{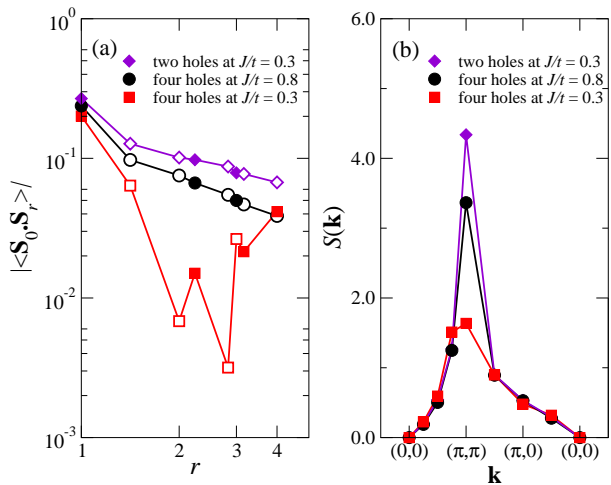}}  
  \caption{\label{fig:ss_sq} (Color online) (a) Spin correlation function
    $\langle{\bf S}_0\cdot{\bf S}_r\rangle$, and (b) static structure factor
    $S({\bf k})$. Empty and filled symbols in (a) represent
    positive and negative correlations respectively.}
 \end{figure}

Up to now we have shown that when the number of holes in a 32-site lattice is
increased from two to four, $d$-wave pairing no longer exists. We note
that at the same time antiferromagnetic (AF) correlation in the spin
background  is also lost.
Evidence comes from
Fig.~\ref{fig:ss_sq} which shows the spin correlation function
$\langle {\bf S}_0\cdot{\bf  S}_r\rangle$ and 
static structure factor $S({\bf k})$. 
At $J/t=0.3$, when we increase the number of holes from two to four,
the spin correlation decreases very rapidly 
and has large fluctuation. 
The sign of the correlation at larger distances shows deviation from
N\'eel order. 
In the momentum space this leads to a strong suppression of the
AF peak in $S({\bf k})$ at $(\pi,\pi)$, leaving no
sign of AF correlation at all. 
Note that this is consistent with the phase diagram of
$\text{La}_{2-x}\text{Sr}_x\text{CuO}_4$, where 
a small doping level of a few percent is enough to destroy the AF phase.
This partially answer the question we posed at the end of the first
paragraph in this paper --- when the doping level is large enough to
destroy AF correlation, holes do not form bound pairs. Instead they become
mutually repulsive.
Then how about at lower doping level where AF correlation still exits?
Do our two-hole results really 
represent the behavior of a lightly doped system, or are they just due
to the artifact 
that there is only one hole pair?
To answer this question directly one has to solve the $t$-$J$ model
with a few hole pairs on a larger lattice. 
But this is currently not feasible. 
In order to mimic a lightly doped
system with a few hole pairs, we put two hole pairs on a 32-site
lattice with an AF spin background.
For this purpose we increase 
$J/t$ to $0.8$. The ground state at this parameter has energy
$-31.917820t$, momentum (0,0) and $s$ symmetry.
Fig.~\ref{fig:ss_sq}(a) shows that its spin correlation decays almost
as a power law, showing that it has at least local antiferromagnetic
order. Its static structure factor clearly has a peak at
$(\pi,\pi)$. We therefore believe that the four-hole system at
$J/t=0.8$ 
possesses AF order. Next we are going to
investigate the possibility of $d$-wave pairing in this system.

\begin{figure}
   \resizebox{8cm}{!}{\includegraphics{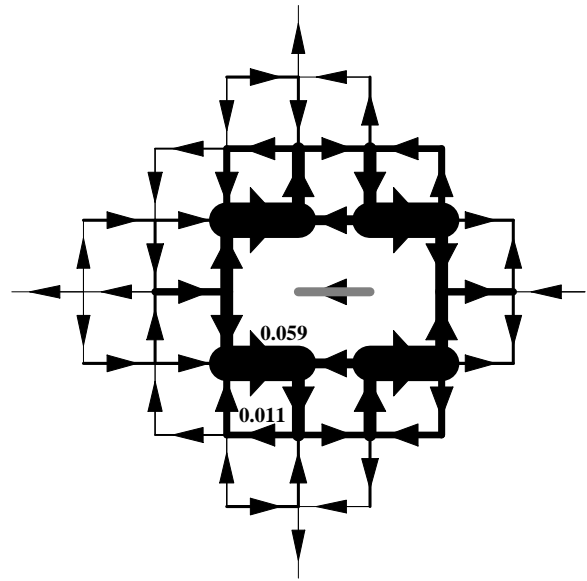}}   
  \caption{\label{fig:jj1} Same as Fig.~\ref{fig:jj} except
   at $J/t=0.8$.}
\end{figure}

\begin{figure}
   \resizebox{6cm}{!}{\includegraphics{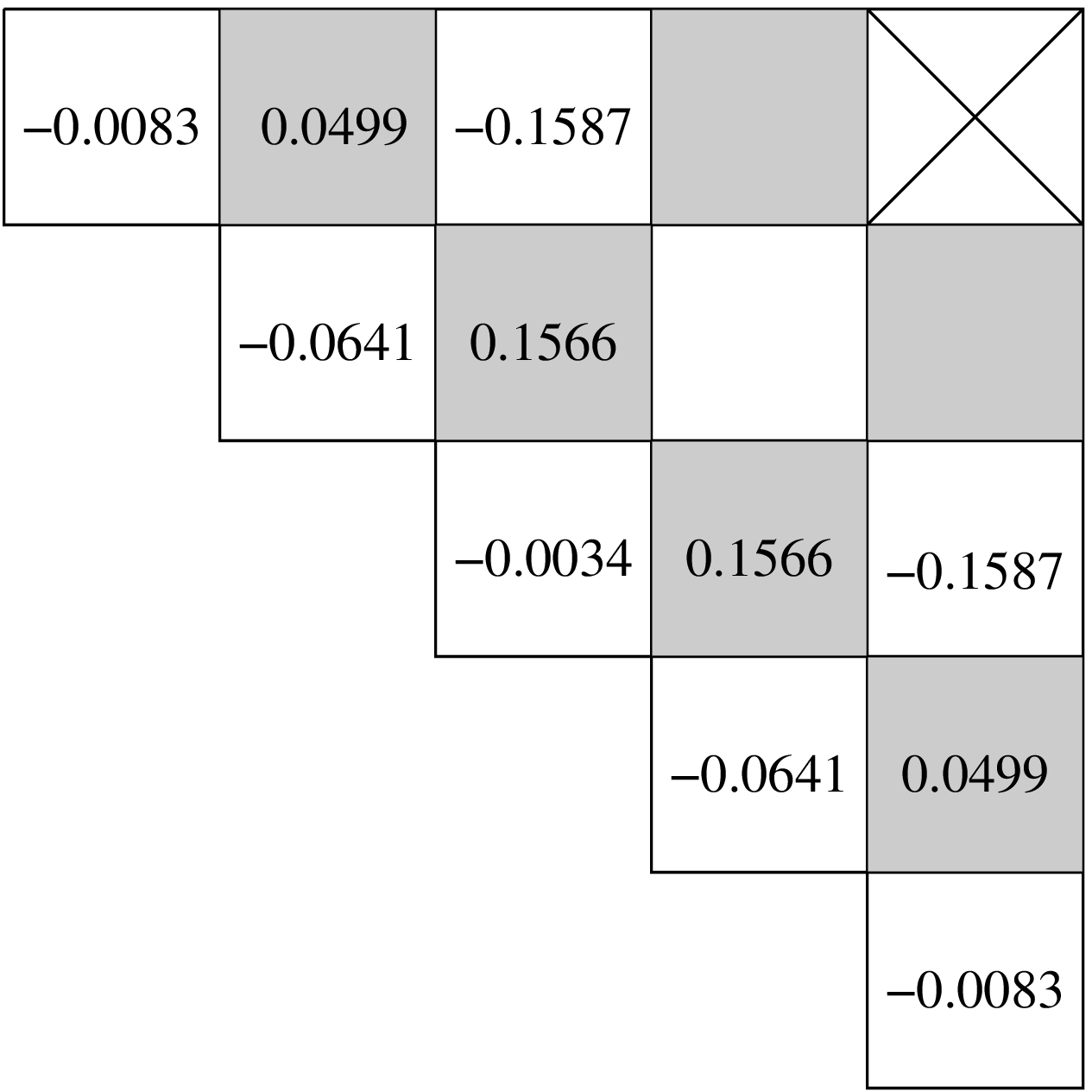}}
  \caption{\label{fig:vv1} Same as Fig.~\ref{fig:vv} except at $J/t=0.8$.}
\end{figure}

A first sign of $d$-wave pairing at $J/t=0.8$ comes from the current
correlation shown in Fig.~\ref{fig:jj1}. Except for a few bonds at or
near the edge of the lattice, the directions of the current correlation
at all other bonds show the correct
staggered pattern. Compared to the result at $J/t=0.3$
(Fig.~\ref{fig:jj}), the current correlation is much stronger in the
present case, especially at small distances.
When we sum up the current correlation around plaquettes to obtain the
vorticity correlation, a clear staggered pattern emerges.
Fig.~\ref{fig:vv1} clearly shows that the vorticity correlation has a
staggered pattern with a characteristic $\pi$ phase shift,\cite{ilw00}
i.e., the sign of $\langle V(r)V(0)\rangle$ is $(-1)^{r_x+r_y+1}$. Note
that the vorticity correlation decays quit fast, which is a
characteristic of strong hole binding at $J/t=0.8$.\cite{l00}
This shows that the four-hole system at $J/t=0.8$ possesses at least
short-range staggered current correlation.
A second sign of $d$-wave pairing at $J/t=0.8$ comes from the hole
correlation $C(r)$ in
Fig.~\ref{fig:cr}. It consists of two regions with different
behavior. At $r\leq 2$ it resembles that of a two-hole system which
shows a decaying trend and with a maximum at $r=\sqrt{2}$. The latter
feature has been shown to be due to the $d$-wave nature of the
two-hole bound state.\cite{we98,rd98} At $r>2$, $C(r)$ remains almost
constant, showing that holes at larger distances are uncorrelated or
even repulsive. This suggests that
holes form $d$-wave bound pairs, and the two hole pairs repel
each other.
Further evidence for this scenario comes from the EMDF. First of all
we note the qualitative similarity between the EMDFs of the four-hole
[Fig.~\ref{fig:h_nq}(b)] and two-hole systems (Fig.~8 of
Ref. \onlinecite{clg98}) --- 
along the AFBZ boundary they are
largest at $(\pi/2,\pi/2)$ and smallest at $(3\pi/4,\pi/4)$.
If the two hole
pairs like to stay away from each other, then their
mutual influence will be a minimum. 
Consequently we expect that the
EMDF of a system with two hole pairs can be built up from that with
one pair in an additive manner.
This is demonstrated in Table~\ref{tab:nq0.8} which shows that
the HMDF of the four-hole system along the AFBZ boundary
is roughly twice that of the two-hole system. 

\begin{table}
\caption{\label{tab:nq0.8} HMDF  in systems with two holes
  ($\langle{\overline n}_{\bf k}\rangle_2$) and four holes
  ($\langle{\overline n}_{\bf k}\rangle_4$) at $J/t=0.8$. {\bf k}
  points along the AFBZ boundary are marked with 
  $\dagger$.}
\begin{ruledtabular}
 \begin{tabular}{ccc}
{\bf k}&
$2\langle{\overline n}_{\bf k}\rangle_2$&
$\langle{\overline n}_{\bf k}\rangle_4$\\
\colrule
 (0,0)                                  & 0.0037    & 0.0028 \\
 ($\frac{\pi}{4}$,$\frac{\pi}{4}$)      & 0.0027 & 0.0046 \\
 ($\frac{\pi}{2}$,$\frac{\pi}{2}$)$^\dagger$      & 0.0728 & 0.0823 \\
 ($\frac{3\pi}{4}$,$\frac{3\pi}{4}$)    & 0.2222 & 0.2196 \\
 ($\pi$,$\pi$)                          & 0.2226 & 0.2179 \\
 ($\pi$,$\frac{\pi}{2}$)                & 0.2463 & 0.2340 \\
 ($\pi$,0)$^\dagger$                              & 0.1395 & 0.1176 \\
 ($\frac{\pi}{2}$,0)                    & 0.0301 & 0.0218 \\
 ($\frac{3\pi}{4}$,$\frac{\pi}{4}$)$^\dagger$     & 0.1498 & 0.1618 \\
\end{tabular}
\end{ruledtabular}
\end{table}


To summarize, our results indicate a close relation between $d$-wave
pairing and AF correlation.
They favor the scenario that
holes in the lightly doped $t$-$J$ model form weak
$d$-wave bound pairs when there is AF correlation in the spin background, and
there is an effective repulsion among the $d$-wave pairs. In this case the
system has all the 
characteristics previously reported for a system with two
holes\cite{clg98,l00,l02} --- the hole correlation function has a 
maximum at $r=\sqrt{2}$, and the EMDF is a
maximum at $(\pi/2,\pi/2)$ along the AFBZ boundary. The current
correlation function 
exhibits a staggered pattern. All these characteristics reflect the
$d$-wave nature of the 
bound state. At higher doping level where AF
correlation disappears, holes no longer form bound pairs and they
become mutually repulsive. This repulsive behavior is very similar to
that of hard core bosons where the repulsion is short range and is due
to lowering of kinetic energy.
The EMDF exhibits dips at $(\pi/2,\pi/2)$
along the 
AFBZ boundary. 
This agrees with mean-field theory predictions,\cite{palee,tk}
although we cannot  
determine whether the dips are hole pockets or segmented Fermi
surfaces.
We emphasize once more that since we cannot study long-range
correlations in our small 
system, we do not know whether those $d$-wave hole pairs can lead to
superconductivity. Therefore although our result indicates the coexistence of
AF and $d$-wave pairing, it {\it does not} necessarily imply the
coexistence of AF 
and superconductivity. 

One popular candidate of a charge-ordered phase in a doped
antiferromagnet is the stripe phase where charge carriers form
one-dimensional strips. It is used to explain the
``1/8 anomaly'' which, incidentally, is at the same doping
level as our four-hole system. 
But since the model we use is isotropic in the $x$-$y$ plane, 
we are not able to detect any
anisotropic correlation. Therefore our results are
inconclusive regarding stripe phase. 
Another possible charge-ordered phase that
can give rise to enhanced charge order at $x=1/8$ is the
checkerboard-type ordering of hole pairs.\cite{chcaz02,kcza04} This
ordering emphasizes the two-dimensional nature of the spin system in
contrary to the one-dimensional nature of the stripe phase.
In the checkerboard-type picture, holes form $d$-wave bound pairs
which in turns form charge-ordered states at a series of magic doping
levels. 
At $x=1/8$ these bound pairs are
arranged in a square lattice of sides 4. 
Incidentally this configuration is compatible with our results 
{\sl when} there
is AF correlation --- the long distance correlation of hole pairs
in the checkerboard-type order
will make hole pairs appear to be repulsive on a small lattice. As a result the
hole correlation will have characteristics of $d$-wave pairs
at small distance
and becomes repulsive at larger distance, very much like the one at $J/t=0.8$
in Fig.~\ref{fig:cr}.
But according to our results
the system must possess at least local AF order in order for this to
be possible.
At $x=1/8$ and realistic value of $J/t$, AF correlation no longer
exists and such 
a charge-ordered phase should not be the ground state.
Nevertheless, this charge-ordered phase may exist as an excited
state. Therefore our results can be 
interpreted as providing indirect evidence for the checkerboard-type
charge ordering.
But we caution that this does not mean that our results favor the
checkerboard-type ordering to stripe phase. Due to its isotropic
nature, the model we use naturally favors two-dimensional ordering to
one-dimensional one
unless other effects (such as lattice distortion) play a role in
selecting the ground state.
Finally we remark that our results do not seem to support phase
separation at 
this doping level because we do not see evidence that four holes tend
to cluster together even when the spin exchange integral $J$ is as
large as $0.8t$.

The author would like to thank T. K. Ng and A. L. Chernyshev for helpful
discussions. 
Calculations on the $t$-$J$ model with four holes were performed on a cluster
of AMD Opteron servers with 64 CPUs.
This work was supported by a grant from the Hong Kong
Research Grants 
Council (Project No. 602004).

Note added -- after this work is finished we learn that
A. Laeuchli has independently solved the $t$-$J$ model with four
holes on a 32-site lattice.


\begin{thebibliography}{}
\bibitem{cr}
$C(r)$ is normalized such that $\sum_{r>0} N_E(r) C(r)=N_h-1$, where
$N_E(r)$ is the number of equivalent site at distance $r$ from the
origin. See 
Eq.(2) of Ref.~\onlinecite{clg98}.
\end{thebibliography}
\end{document}